\documentclass[aps,prb,twocolumn,superscriptaddress]{revtex4-1}

\usepackage{graphicx}
\usepackage{bm}
\usepackage{amsmath}
\usepackage{amssymb}
\usepackage{hyperref}
\usepackage{mhchem}
\usepackage{color}

\begin{document}

\title{Catalytic role of HI in the interstellar synthesis of complex organic molecule}

\author{Shuming Yang}
\affiliation{Laboratory for Relativistic Astrophysics, Department of Physics, Guangxi University, Nanning 530004, China}

\author{Peng Xie}
\affiliation{School of Chemistry and Chemical Engineering, Guangxi University, Nanning 530004, China}

\author{Enwei Liang}
\affiliation{Laboratory for Relativistic Astrophysics, Department of Physics, Guangxi University, Nanning 530004, China}

\author{Zhao Wang}
\email{zw@gxu.edu.cn}
\affiliation{Laboratory for Relativistic Astrophysics, Department of Physics, Guangxi University, Nanning 530004, China}

\begin{abstract}
Using quantum chemical calculations, we model the pathways for synthesizing two purine nucleobases, adenine and guanine, in the gas-phase interstellar environment, surrounded by neutral atomic hydrogen (HI). HI is found active in facilitating a series of fundamental proton transfer processes of organic synthesis, including bond formation, cyclization, dehydrogenation, and H migration. The reactive potential barriers were significantly reduced in the alternative pathways created by HI, leading to a remarkable increase in the reaction rate. The presence of HI also lowered the reactive activation temperature from 757.8 K to 131.5-147.0 K, indicating the thermodynamic feasibility of these pathways in star-forming regions where some of the reactants have been astronomically detected. Our findings suggest that HI may serve as an effective catalyst for interstellar organic synthesis.
\end{abstract}


\maketitle

\section{Introduction}          
\label{sect:intro}

Finding the synthesis pathways of interstellar complex organic molecules (COMs) of biological interest is paramount for explaining the origin of the elementary building blocks of life detected on meteorites, comets and asteroids \citep{Burton2012,Morse2019,Singh2013,Misra2017}. The formation of biomolecules through direct collisions between abiotic molecules is generally challenging in the cold environment of the interstellar medium (ISM). Photochemical and radiation chemical processes, particularly on ice/dust grains, are commonly regarded as the principal steps towards synthesizing interstellar COMs \citep{Materese2020, Oberg2016,Herbst2009,Mifsud2021,Carrascosa2021}. While the ion-neutral or radical routes are the commonly considered mechanisms for the gas-phase synthesis of interstellar COMs, catalyzed reactions are emerging as alternative processes since the majority of the detected interstellar molecules are neutral and have a closed shell.

Catalysts, often in the form of free radicals, ions, or metal clusters, are frequently utilized in laboratory experiments of organic chemistry to convert a challenging molecule-molecule reaction into a more straightforward molecule-radical or radical-radical reaction \citep{Qin2017,Zhao2020}. For the synthesis of interstellar organics, previous works suggest that ice/dust grains, hydronium, and carbamic acid could function as catalysts \citep{Potapov2019,Roy2007,Silva2017,Saladino2013,Qi2018,Hanine2020,Mendoza2004,Rimola2012,Vinogradoff2015}. For example, radicals such as ·\ce{NH2} and ·CN are considered to participate the reactions for synthesizing nucleobases from HCN, and are capable of reducing the barrier and may even render some reactions barrierless \citep{Jeilani2013,Jeilani2016}. Nonetheless, the low abundance of the previously suggested catalysts often renders the catalyzed reaction seemingly impossible in ISM, as it requires the collision of three bodies. Fortunately, this is not always the case. If the concentration of the catalyst is in great excess of that of the reactants, the reaction can occur just as easily as a two-body collision. Neutral atomic hydrogen (HI) is the simplest and the most abundant free radical in the universe. It dominates the chemical composition of the ISM, and is virtually ubiquitous in every interstellar environment. In hot molecular cores, the concentration of HI is about $7-10$ orders of magnitude higher than that of COMs \cite{Herbst2009}.  

Although HI is a common reactant or product in simple interstellar reactions, its catalytic activity in interstellar organic synthesis remains unclear in most astrochemistry models  \citep{Bergantini2014,Carrascosa2020,Chuang2016,Fedoseev2015,Fedoseev2016,Miksch2021,Slate2020,Singh2014,Misra2017}. In comparison to its molecular form, atomic hydrogen is rare on Earth and is difficult to prepare and control in laboratory experiments. As a result, quantum chemical calculations are expected to shed light on whether HI plays a crucial role in the interstellar formation of COMs \citep{Zamirri2019}. Aiming at this question, we conducted a new series of ab initio quantum chemical calculations to explore the potential catalytic activities of HI in the formation pathways of adenine (Ade) and guanine (Gua).

Ade and Gua are two purine nucleobases constituting the genetic code in the nucleic acids. After having been detected in meteorites, their extraterrestrial origin and prebiotic formation are a long-debated focus of attention in attempts to answer the abiogenesis question \citep{Burton2012,Peeters2003}. Intensive efforts are therefore devoted to find the synthesis pathways of nucleobases in laboratory experiments simulating the chemical evolution of abiotic molecules in interstellar environments \citep{Sandford2020,Chuang2016,Fedoseev2016,Parker2017,Merz2014,Bera2017,Glaser2007,Choe2021,Silva2017,Parker2015,Gupta2011,Fulvio2021,Parker2015}. In particular, Ade, Gua and purine derivatives have been synthesized through photochemical reactions of purine in an interstellar ice analogue \citep{Materese2017,Materese2018}. 2-aminopurine and isoguanine are assumed to be two key intermediates for these reactions. Nevertheless, the HI-catalyzed synthesis of nucleobases has yet to be investigated.

\section{Methods}
\label{sect:Obs}

\begin{figure}[htbp]
\centering
\includegraphics[width=9cm]{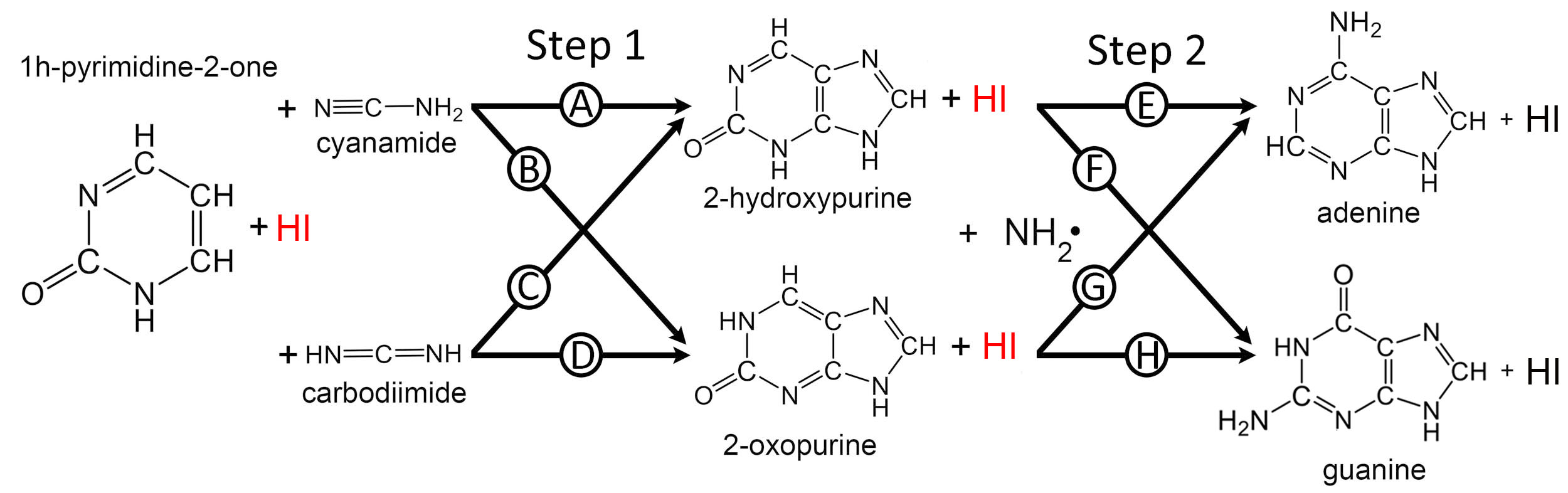}
\caption{Formation scheme of Ade and Gua.} 
\label{F1}
\end{figure}

We consider novel synthesis pathways of Ade and Gua starting from 1h-pyrimidine-2-one \ce{C4H4N2O}, a direct intermediate toward nucleobases producted by two interstellar molecules: formamide \ce{H2NCHO} and vinyl cyanide \ce{H2CCHCN} \citep{Lu2021}. Our proposed pathways involve two steps, illustrated in Figure~\ref{F1}. In the 1st step (via the paths A-D), \ce{C4H4N2O} reacts with either cyanamide (\ce{NH2CN}) or carbodiimide (\ce{HNCNH}) (both detected in Sgr B2 and Orion-KL \citep{Turner1975,McGuire2012,Belloche2013,White2003} to form two direct purine-base precursors, 2-hydroxypurine (\ce{C5H4N4O}) and 2-oxopurine (\ce{C5H4N4O}), via the paths A-D. In the 2nd step, these two precursors produce Ade or Gua by separately reacting with ·\ce{NH2} via the paths E-H. We focus on the catalytic activities of HI in the both two steps. 

Multiple possible collision trajectories were simulated for each reaction pathway. The transitional and final molecular structures in single molecular collision events were determined by structural optimization using first principles calculations \citep{Maeda2015,Barone2015,Puzzarini2022,Zamirri2019,pan2021}, in which the system's total potential is minimized in the framework of the density functional theory (DFT) at M06-2X/6-31+G(d,p) level as implemented in Gaussian 16 B.01 \citep{Frisch2016}. We chose this level of theory in view of its proven records of accuracy in predicting physical chemical properties of relevant compounds \citep{Mardirossian2017,De2021}. In our previous work \cite{Lu2021}, this method was compared with calculations based on M\o ller Plesset perturbation theory in the second order (MP2), and a good agreement was achieved. 

Vibration frequency calculations were here carried out to confirm the intermediate (IM) and the transition state (TS). Intrinsic reaction coordinate (IRC) calculations were performed to ensure that the TS reasonably connect the reactant (RC) and the product (PD). To incorporate HI in our calculations, we initially positioned a HI in a transition state with the reactant. We then performed IRC coordinate scanning to ensure that the HI molecule's placement was appropriate. Subsequently, we simulated multiple trajectories of collisions between the HI and the reactant from various positions. A static calculation is used for this step, unlike what occurs in molecular dynamics \cite{wang2020,wang2019}. Finally, we selected trajectories with the lowest energy cost to facilitate the reaction. The atomic coordinate data of all optimized molecular structures, including RC, IM, TS, and PD, are provided in the standard XYZ chemical file format in the Supplementary Information's Data section.

The Gibbs free energy $G$ was calculated for every optimized structure of RC, IM, TS or PD at a thermodynamic temperature of $150$ K under the assumption of non-interacting particles \citep{Mcquarrie1999} (p321-325). This representative temperature corresponds to a typical temperature of hot molecular cores. The potential barrier $\Delta G$ of a reaction pathway was determined as the difference in $G$ between the highest-energy states (including IM, TS, or PD) and the RC in each of the two steps described in Figure \ref{F1}. This definition of the reactive barrier assumes that the energy does not leave the system by irradiation on a timescales typical of the atomic rearrangements, which is in general much faster than the needed vibration cascade. As multiple collision trajectories were simulated for each path, only the energetically favorable trajectories with the lowest $\Delta G$ are discussed in the following. 

\section{Results and discussions}

\begin{figure}
\centering
\includegraphics[width=9cm]{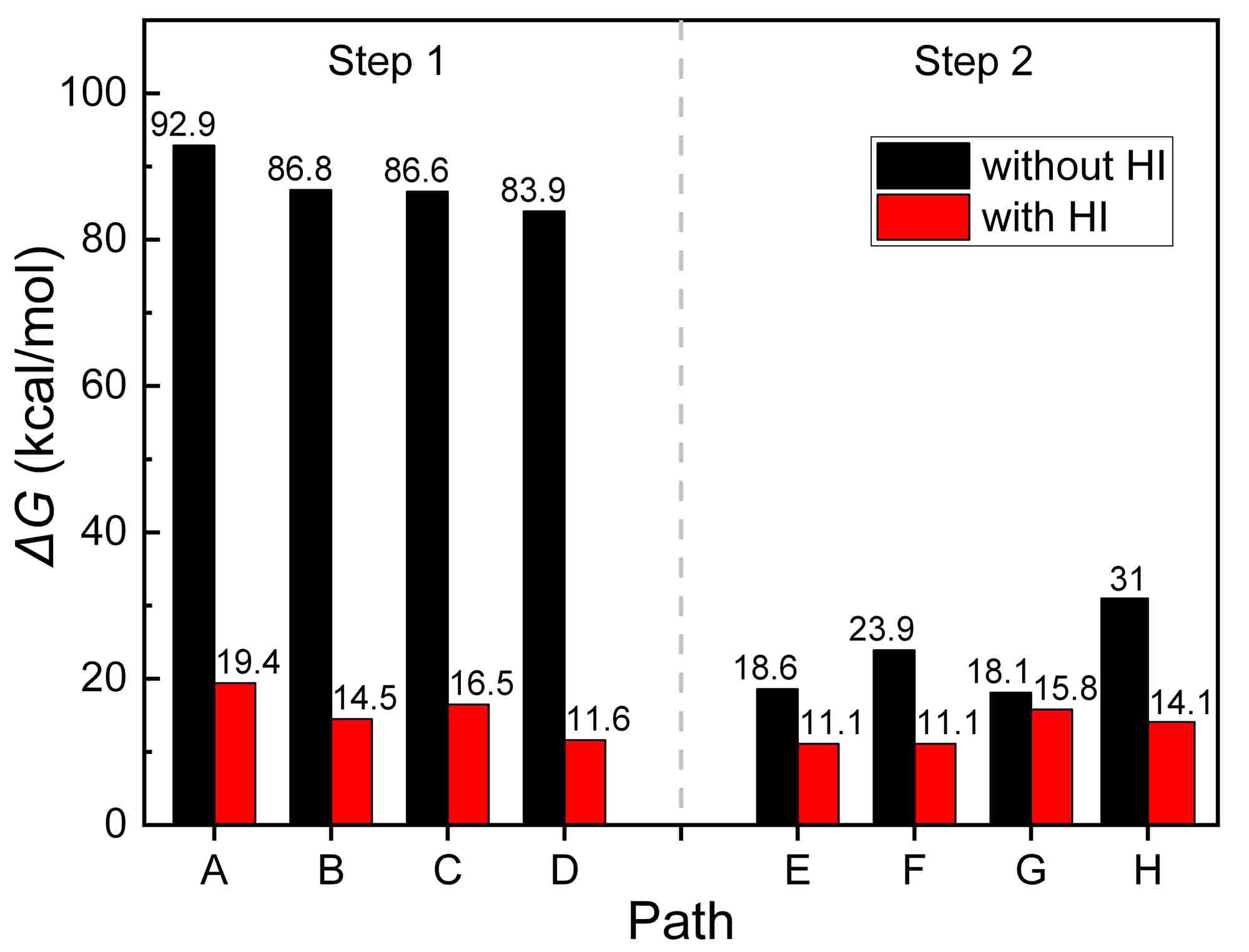}
\caption{Minimum reactive potential barrier for different reaction paths.} 
\label{F2}
\end{figure}

the reactive barrier $\Delta G$ is a crucial parameter for assessing the feasibility and significance of a reaction in the ISM, as it strongly correlates with the reaction rate and activation temperature. Figure~\ref{F2} displays the minimum energy costs of producing Ade and Gua from \ce{C4H4N2O} through various paths, and compares $\Delta G$ for reactions with (red bars) and without (black bars) the involvement of HI. The results demonstrate that HI greatly promotes the reactions.

In the absence of HI, the reaction pathway (Path A-D, left half) shows a high barrier with $\Delta G > 83$ kcal/mol, which is expected since the fused pyrimidine-imidazole rings in Ade and Gua are complex to synthesize without a catalyst. However, the addition of HI significantly facilitates the reactions, reducing $\Delta G$ to $11.6-19.4$ kcal/mol. Path D, involving \ce{HNCNH}, shows the lowest barrier among the proposed paths, indicating that it is more active than its isomer \ce{NH2CN}. Considering steps 1 and 2 consecutively, the optimal pathway is either Paths B+G/D+G or D+H, with HI as a catalyst. The corresponding optimal barrier for synthesizing Ade or Gua is around $15.8$ or $14.1$ kcal/mol, respectively.

\begin{figure}
\centering
\includegraphics[width=9cm]{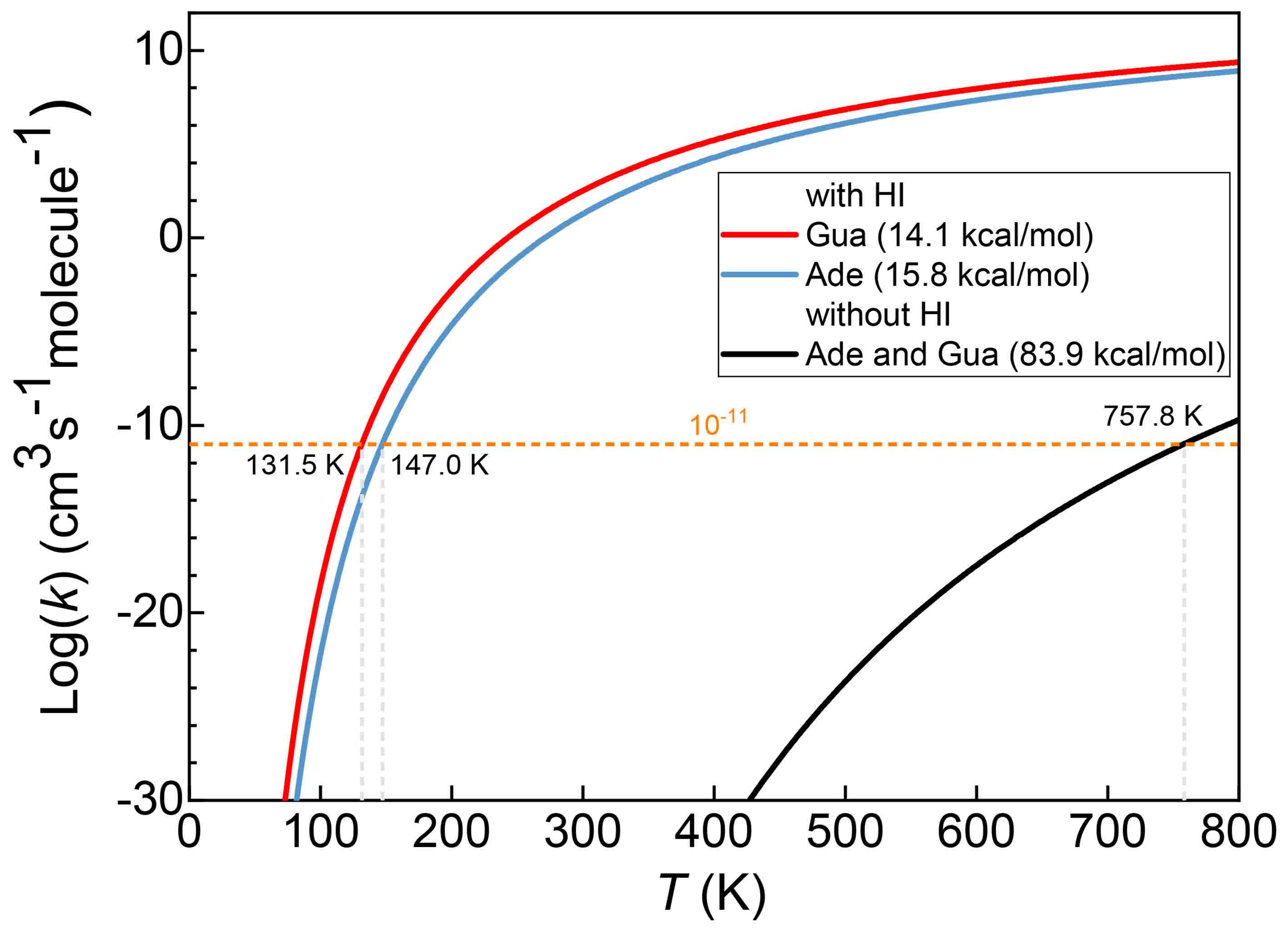}
\caption{Reaction rate constant versus temperature for the optimal paths without (black curve) and with HI (red and blue curves). The ternary reactions with HI are assumed to be pseudo binary reactions.} 
\label{F3}
\end{figure}

As the temperature of a gas rises, its constituent molecules collide more frequently and with greater kinetic energy. As a result, the fraction of collisions that can surmount the barrier $\Delta G$ increases, following the Arrhenius relation. In chemical kinetics, the Eyring-Polanyi equation \citep{eyring1935} describes the relationship between the reaction rate constant $k$, $\Delta G$, and $T$,

\begin{equation}
k=\frac{C \kappa_{B} T}{h}e^{-\frac{\Delta G}{RT}}. 
\label{Eq1}
\end{equation}

\noindent where $C$ is the transmission coefficient, $\kappa_{B}$ is the Boltzmann constant, $R$ is the molar gas constant, and $h$ is the Planck constant. Using this equation, we estimate $k$ as a function of $T$ as plotted in Figure~\ref{F3}, for the optimal reactive paths of Ade and Gua with and without HI. Generally speaking, most significant reactions in the ISM typically proceed with a rate constant of 10$^{-11}$ cm$^{3}$s$^{-1}$ or larger (above the dashed horizontal orange line). Without catalyst, such a rate constant would roughly require a reactive temperature $>757.8$ K with $\Delta G = 83.9$ kcal/mol, as indicated by the dashed line intersecting with the black curve in Figure~\ref{F3}. The feasibility of such a reaction would be low in the generally cold, gas-phase environment of dense molecular clouds, as the reactants would have to coincide with radiation having an appropriate amount of energy at the time of collision. In contrast, the presence of HI significantly accelerates the reaction, leading to a substantial increase in the reaction rate constant by approximately $20-30$ orders of magnitude (as seen in Figure~\ref{F3}). As a result, the required activation temperature is respectively reduced to only $131.5$ or $147.0$ K for Gua or Ade, as indicated by the dashed lines intersecting with the red and blue curves. To verify our computational model, we compute the half-life time based on the reaction rate and compare it to the results of \citet{Zamirri2019}, with agreement obtained as shown in Part I of the Supplementary Information.

Three-body collision processes are often deemed impossible in the ISM, but this does not hold true for the HI-catalyzed reaction. This is because the concentration of HI vastly exceeds that of COMs, reaching 10$^{7}$-10$^{10}$ times in hot cores \cite{Herbst2009}. Thus, ternary reactions with HI are essentially pseudo-binary reactions. In chemistry models for star-forming regions, the HI abundance is estimated to be comparable to that of H2 \citep{Garrod2008,Sun2022}. In addition to the remaining background HI from large-scale molecular clouds, HI also stems from the photo-dissociation of H2 by the radiation emitted from YSOs. This in particular points to the inner parts of protoplanetary disks, in which the ultraviolet radiation can be dramatically enhanced and resulting in a chemical structure resembling that of a photon-dominated region (PDR) \citep{van2006}. In such a region, photons are usually not energetic enough to ionize hydrogen but can dissociate molecular hydrogen \citep{Hollenbach1999}. For example, the total HI mass in a protoplanetary disk is estimated to be $11.4$ $M_{Earth}$ with a typical vertical column density of $1.7 \times 10^{21}$ or $3.3 \times 10^{20}$ cm$^{-2}$ at $5$ or $110$ AU, respectively \citep{kamp2008}. Moreover, HI is a common product of many important reactions in ISM \citep{Herbst1973}, as proton transfer is a fundamental process in organic reaction \citep{Mayer2011}.

The involvement of HI makes the proposed pathways thermally feasible in certain interstellar environments with low activation temperatures. For instance, the formation sites of young stellar objects (YSO) are often associated with warm ambient gas, known as hot cores or corinos, with average temperatures up to $300$ K, where a rich organic chemistry is observed  \citep{Herbst2009,Jorgensen2020}. Interestingly, most of here-used reactants, including formamide (\ce{NH2CHO}), vinyl cyanide (\ce{CH2CHCN}), \ce{NH2CN}, and \ce{HNCNH}, are detected to co-exist in SgrB2 and Orion KL hot cores \citep{Turner1975,McGuire2012,Belloche2013,White2003,Scibelli2021}. Note that the reactant \ce{C4H4N2O} is supposed to be synthesized from \ce{NH2CHO} and \ce{CH2CHCN} \citep{Lu2021}, which are also both detected in those regions as ``hot'' molecules with characteristic rotation temperature $>100$ K \citep{Lopez-Sepulcre2019,Belloche2013,Lopez2014,Suzuki2018,Fu2016}. Those regions contain the initial materials as well as enough thermal energy for the proposed reactions to proceed. It is noteworthy that protoplanetary disks are also of particular interest for abiogenesis, as the planets formed in them inherit their chemical compositions. Although only simple organics have been detected in protoplanetary disks so far \citep{Oberg2020}, the activation temperature reported in this study for nucleobase synthesis is lower than that in the inner part of disks, as the temperature gradually increases from $10$ K to several thousand K from the outer layer to the central forming star \citep{Akimkin2013}.

\begin{figure}
\centering
\includegraphics[width=9cm]{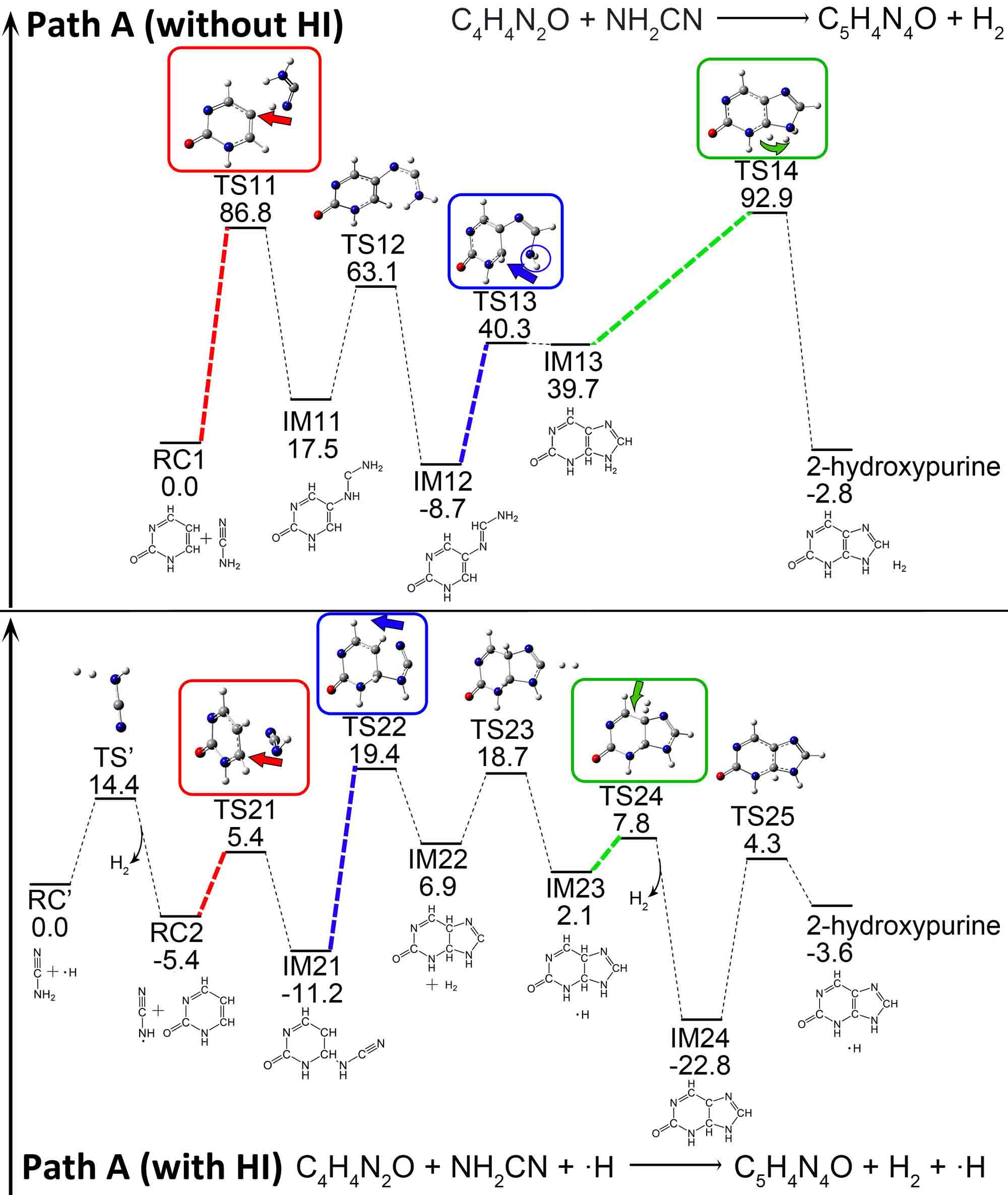}
\caption{Potential energy surface for forming 2-hydroxypurine (\ce{C5H4N4O}) from \ce{C4H4N2O} and \ce{NH2CN} in via Path A without (upper panel) or with HI (lower panel). The arrows indicate the direction of molecular collision.} 
\label{F4}
\end{figure}

One might naturally inquire about the specific effects of HI in promoting the reaction. To showcase the catalytic capabilities of HI, Figure~\ref{F4} provides a comparison of the reaction pathways in the absence and presence of HI, using path A as an example. HI is found to facilitate several critical processes, including cyclization, dehydrogenation, and the bonding of the reactants. In the absence of HI, the reaction is impeded by a significant barrier of $86.8$ kcal/mol against the formation of an initial bond between the two reactants (see the step in the red line, upper panel). However, when HI is present in the vicinity, it can extract a hydrogen atom from the amino group with a relatively low cost of only $14.4$ kcal/mol, converting \ce{NH2CN} into the highly-reactive HNCN radical. This generates an alternative pathway in which the barrier for bonding the two reactants is considerably lowered to $10.8$ kcal/mol, as indicated by the red line in the lower panel.

In path A, the rate determining step involves two conventional organic synthesis processes, namely cyclization and dehydrogenation. Initially, cyclization exhibits a high barrier of around $49.0$ kcal/mol, as indicated by the blue circle in the upper panel of Figure~\ref{F4}. This barrier arises because the N end required for forming the imidazole ring is completely saturated. However, this process is expedited in the alternative pathway created by HI, where the (1h-pyrimidine-2-one-6-yl)-cyanamide radical (IM21) possesses an unsaturated N end, as depicted by the blue rectangle in the lower panel of Figure~\ref{F4}. Although cyclization still requires approximately $30.6$ kcal/mol, the energy retained in the preceding sub-steps helps to surmount this barrier, owing to the low energy of IM21.

The next step is dehydrogenation, in which a C-bonded and an N-bonded hydrogen atom in the imidazole ring are converted into a H2 molecule, resulting in the formation of 2-hydroxypurine \ce{C5H4N4O}, as shown in Figure~\ref{F4}. The original barrier for this process is roughly $53.2$ kcal/mol, which is substantially reduced in the alternative pathway created by HI in the lower panel (green line). One of the H atoms in IM23 is taken by the free HI, at a small cost of only $5.7$ kcal/mol, to generate the stable IM24, 1,6-dihydro-purin-2-one radical. This radical ultimately yields 2-hydroxypurine (\ce{C5H4N4O}) by directly removing another H atom.

\begin{figure}
\centering
\includegraphics[width=9cm]{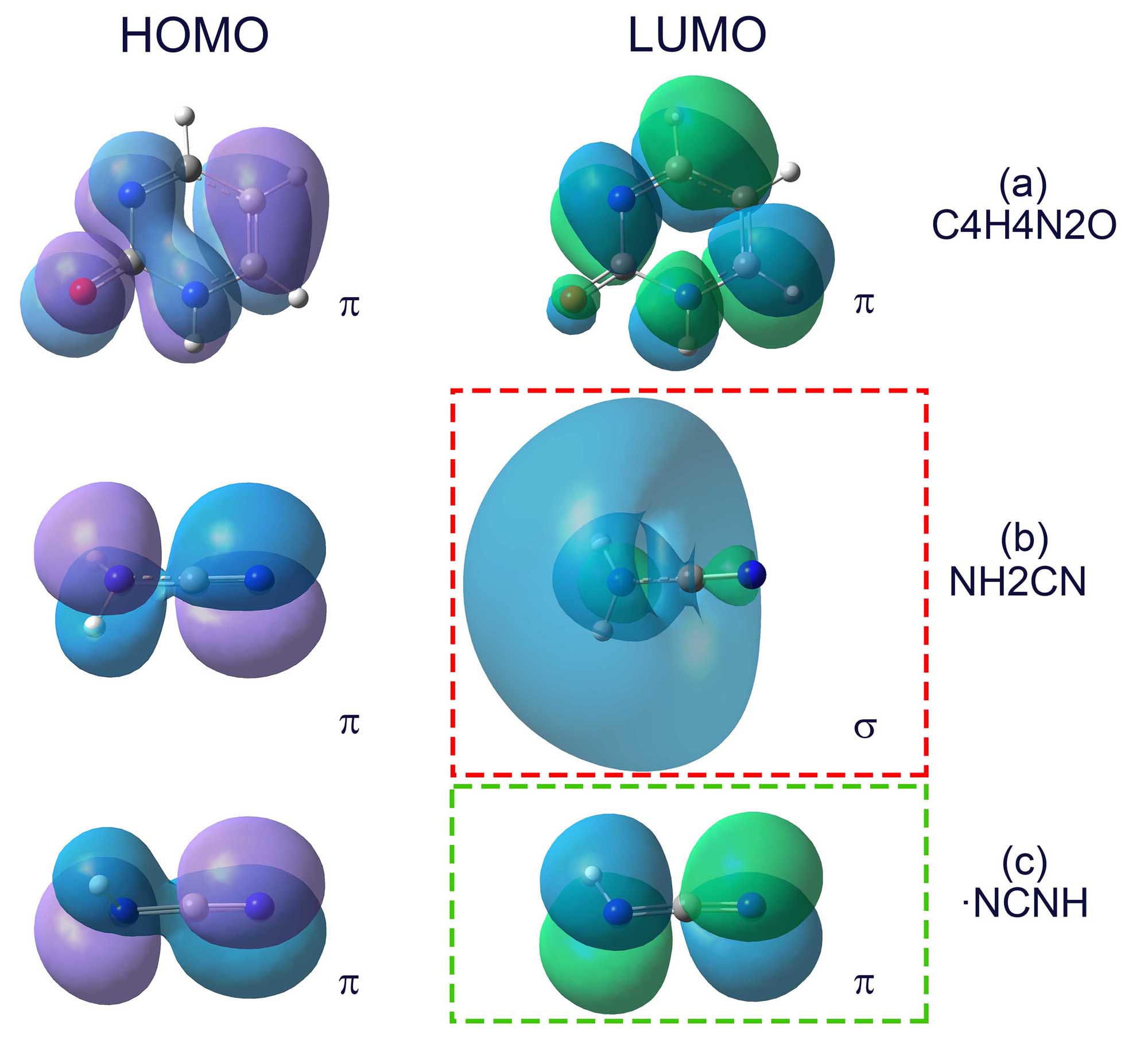}
\caption{HOMO (left) and LUMO (right) of C4H4N2O (a), NH2CN (b) ,HNCN (c).} 
\label{F5}
\end{figure}

The catalytic activity of HI is demonstrated using the initial step of Path A in Figure~\ref{F4} as an example within the framework of the frontier molecular orbital theory. This step involves the formation of a bond between the reactants, \ce{C4H4N2O} and \ce{NH2CN}, to initiate the reaction. However, because the highest occupied molecular orbital (HOMO) of \ce{C4H4N2O} (located on the left in Figure~\ref{F5} (a)) and the lowest unoccupied molecular orbital (LUMO) of NH2CN (outlined in red) have mismatched symmetry ($\pi$ vs $\sigma$ orbital), a high energy is required to form the bond. The feasibility of a reaction process is closely related to the conservation of orbital symmetry, and a lack of symmetry can impede the reaction \citep{woodward1965,woodward1969}. Consequently, a significant HOMO-LUMO gap of $7.1$ kcal/mol arises, making the reaction difficult to proceed. Fortunately, HI has the ability to extract a hydrogen atom from \ce{NH2CN}, thereby transforming it into a radical \ce{HNCN}, which can readily initiate the reaction because its LUMO symmetry (outlined in green) is consistent with the HOMO symmetry of \ce{C4H4N2O} ($\pi$ vs $\pi$ orbital). As a result, the HOMO-LUMO gap is significantly reduced to $3.6$ kcal/mol, allowing for the reaction to proceed smoothly.

\begin{figure}
\centering
\includegraphics[width=9cm]{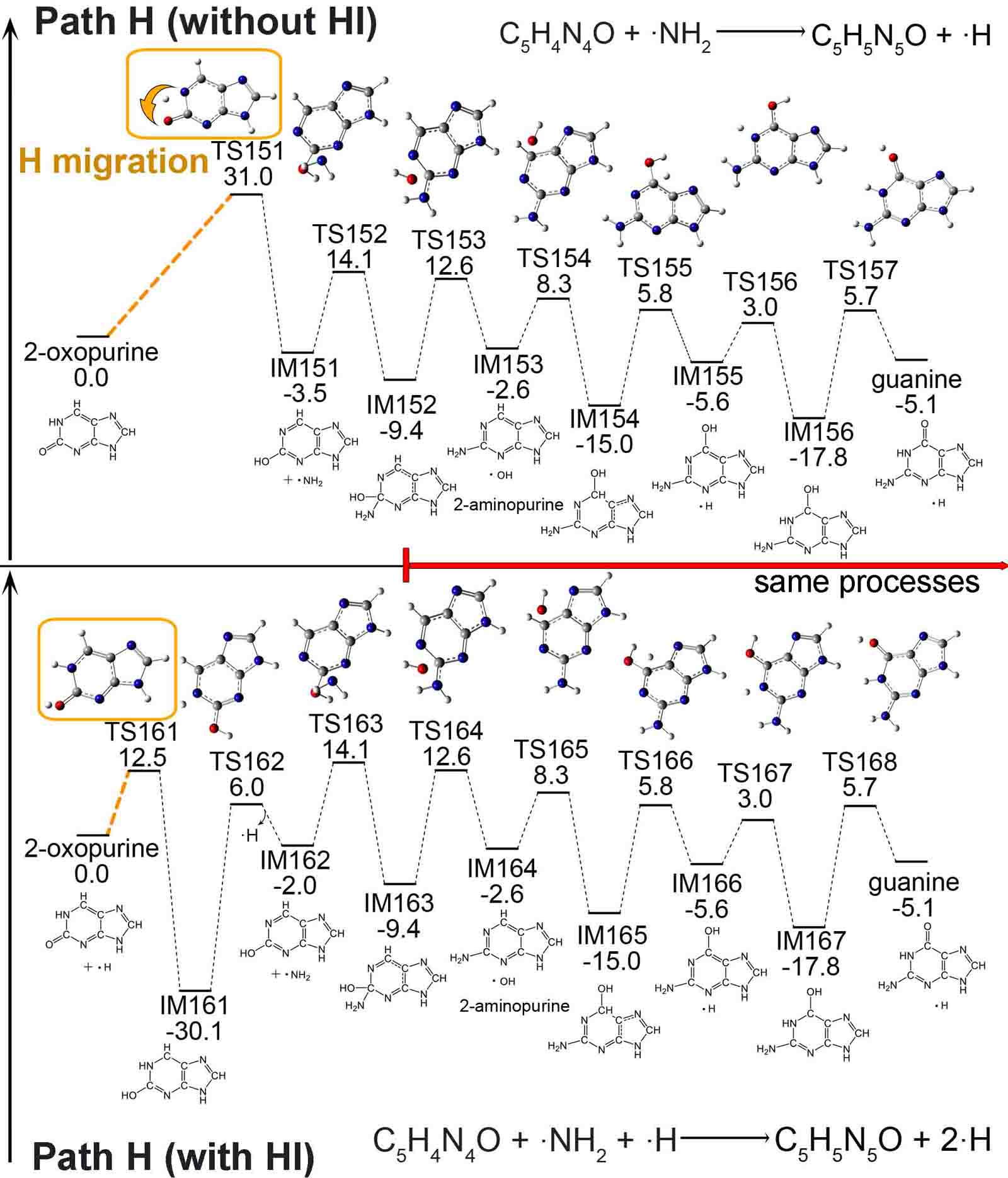}
\caption{Potential energy surface to form Gua from 2-oxopurine (\ce{C5H4N4O}) and ·\ce{NH2} via Path H without (upper) or with HI (lower). The processes in the two paths become the same after IM152 and IM163.} 
\label{F6}
\end{figure}

HI has also been found to aid in hydrogen migration, as demonstrated in path H, where 2-oxopurine (\ce{C5H4N4O}) reacts with ·\ce{NH2} to form Gua, as depicted in Figure~\ref{F6}. To initiate this reaction, a proton must move from an N site to an adjacent O site, which requires approximately $31.0$ kcal/mol (see yellow line in the upper panel). This barrier can be reduced by the presence of HI, which binds to the relevant O atom, creating a highly stable intermediate, IM161, 2-hydroxy-3H-Purine radical, as shown in the lower panel. This intermediate sits in a deep potential well, and the energy it contains aids in overcoming the subsequent processes of dehydrogenation and bonding between ·\ce{NH2} and a $sp^{2}$-hybridized C.

The catalytic effects of HI are observed in all of the simulated reaction paths (B-G), as provided in Part II of the Supplementary Information. Additionally, several stable intermediates are identified in the proposed reaction pathways, including 2-aminopurine, purinol, 2-hydroxypurine, 2-oxopurine, and isoguanine. Some of these intermediates align with the findings of prior laboratory experiments. For instance, 2-aminopurine (IM153 or IM164 in Figure~\ref{F6}) was discovered by \citet{Materese2017,Materese2018} in a photochemistry experiment on an interstellar ice analogue and was believed to be a critical intermediate in the synthesis of Ade, Gua, and purine derivatives. The results of this experiment also suggest the same function for isoguanine and purinol. The present study elucidates the pathways from these intermediates to the purine bases. Finally, the IR and electronic adsorption spectra of these intermediates are calculated using DFT calculations \citep{Kovacs2020,Meng2021,wu2022} and are presented in Part III of the Supplementary Information to facilitate potential astronomical observations.

\section{Conclusions}
\label{sect:discussion}

The proposed new pathways for the synthesis of Ade and Gua demonstrate the efficiency of HI as a catalyst in promoting various fundamental organic synthesis processes, including bond formation, cyclization, dehydrogenation, and H migration. These are typical proton transfer processes in organic synthesis, where HI can effectively lower the potential barrier and significantly increase the reaction rate constant by up to $20-30$ orders of magnitude, depending on the temperature. As a result, the activation temperature required for the proposed reaction is lowered from 757.8 K to 131.5-147.0 K when HI is present. This suggests that the thermal feasibility of gas-phase purine base synthesis with HI could be possible in environments such as hot molecular cores or protoplanetary disks. From a broader perspective, given the prevalence of HI as a major component of interstellar gas and the fundamental role of proton transfer in organic synthesis, the HI-catalyst mechanism should have wide implications for the formation of organic molecules in the ISM.

\end{document}